\documentclass[final,5p,times,twocolumn]{elsarticle}

\journal{Physics Letters B}

\begin{document}

\begin{frontmatter}



\title{Temporal evolution of tubular initial conditions and their
influence  on two-particle correlations in relativistic nuclear collisions}


\author[label1]{R.P.G.Andrade}
\author[label1]{F.Grassi} 
\author[label1]{Y.Hama}
\author[label2]{W.-L.Qian}

\address[label1]{Instituto de F\'{\i}sica, Universidade de S\~ao Paulo, Brazil}
\address[label2]{Instituto de Ci\^encias Exatas, Universidade Federal de Ouro 
Preto, Brazil}

\begin{abstract}
Relativistic nuclear collisions data on two-particle correlations 
exhibit structures as function of
 relative azimuthal angle and rapidity. 
A {\em unified} description of these near-side and away-side structures is 
proposed for low to moderate transverse momentum.
It is based on the combined effect of tubular initial conditions and 
hydrodynamical expansion. Contrary to expectations, the hydrodynamics 
solution shows that the
high energy density tubes (leftover from the initial particle interactions) 
give rise to particle emission in two 
directions and this is what leads to the various structures.
This description is sensitive to some of the initial tube parameters and may
provide a probe of the strong interaction. 
This  explanation is compared with  an alternative one where some 
triangularity in the initial conditions is assumed.
A possible experimental test is suggested.

\end{abstract}


\end{frontmatter}


\section{The need for a unified description}

One of the most striking results in relativistic heavy-ion collisions 
at RHIC and the LHC,
is the 
existence of  structures
  in the two-particle correlations \cite{R2,star3,R3,R1,phoboss,Ralice,Rcms,Ratlas} plotted as function of the pseudorapidity difference $\Delta \eta$ and 
the angular spacing $\Delta \phi$.
The so-called ridge has a narrow $\Delta \phi$ located around zero and a
long $\Delta \eta$   extent.
The other structure
located opposite has a single or 
double hump in     $\Delta \phi$.
In order that two particles,
emitted at some proper
 time $\tau_{f.out}$, appear as correlated over several rapidity units,  
the process that  correlated
them must have occurred  \cite{lml1,lml2} at a much smaller proper time
due to causality. 
Therefore, the existence of  long range pseudorapidity correlations 
 must be related to early times in the nuclear collisions and thus 
has motivated many theoretical 
investigations.

Hydrodynamics has now been established as a good tool to describe many
data from relativistic heavy-ion collisions so it should be able to provide 
a 
description for the above mentioned structures
(for low to intermediate
transverse momenta).
In fact, as noted with RHIC data, a 
 hydrodynamics based explanation is attractive because of the various
similarities (see e.g. \cite{STARbulk})
between bulk matter and ridge (transverse momentum spectra,
baryon/meson ratio, etc). 
In addition, it was shown (particularly at the LHC) that particle correlations
can be understood in term of anisotropic flow Fourier coefficients
\cite{ALICEfact,ATLASfact,CMSfact}.
This points towards the necessity to have a {\em unified} 
hydrodynamic description
of near and away-side structures.

In early models, it was suggested that the combined effect of
longitudinal high energy density tubes 
(leftover from initial particle collisions)   and 
transverse expansion was 
responsible for the ridge \cite{voloshin,shuryak,lml1,lml2,sg}.
The particle emission associated to the tube was expected to occur in a single
direction, so this would cause a ridge but no
 away-side structure. In addition,
 the effect of 
hydrodynamics was usually assumed to be of a certain type (e.g. a blast wave in
 \cite{lml1,lml2,sg})
and in fact when hydrodynamic expansion was actually computed, it
seemed to lead to a disappearance 
of the initial high energy
density tubes \cite{shuryakhole} and therefore of their particle emission.

In a previous work \cite{jun}, we presented evidence that
hydrodynamics might in fact reproduce all structures 
using the  NeXSPheRIO code. This code starts with
initial conditions from the event generator NeXus \cite{nexus} and solves the hydrodynamic equations on an 
event-by-event basis with the method of Smoothed Particle 
Hydrodynamics. In \cite{jun}, the NeXSPheRIO events were 
analyzed in a similar way to the experimental ones, in 
particular the ZYAM method was used to remove effects of 
elliptic flow. We later developed a different method to 
remove elliptic flow from our data and  checked that all 
structures were indeed exhibited and other features well 
reproduced (dependence on the trigger- or associated-particle 
transverse momentum, centrality, in-plane/out-of-plane trigger,
 appearance of a peak on the ridge). 
However, when using NeXSPheRIO, it is not clear how the 
various structures in the two particle correlations are 
generated. The aim of this paper is to investigate this,  studying 
in detail what happens in the vicinity of an energetic tube (section 2) and 
then extending the results to a more realistic complex case (section 3). 
We will also compare our explanation with 
an alternative one that assumes some triangularity of the initial
conditions (section 4).

\section{A simplified model}

\subsection{Origin of the near-side and away-side structures}

\begin{figure*}[htb]   
\begin{minipage}[h]{120mm} 
\includegraphics[width=6.cm]{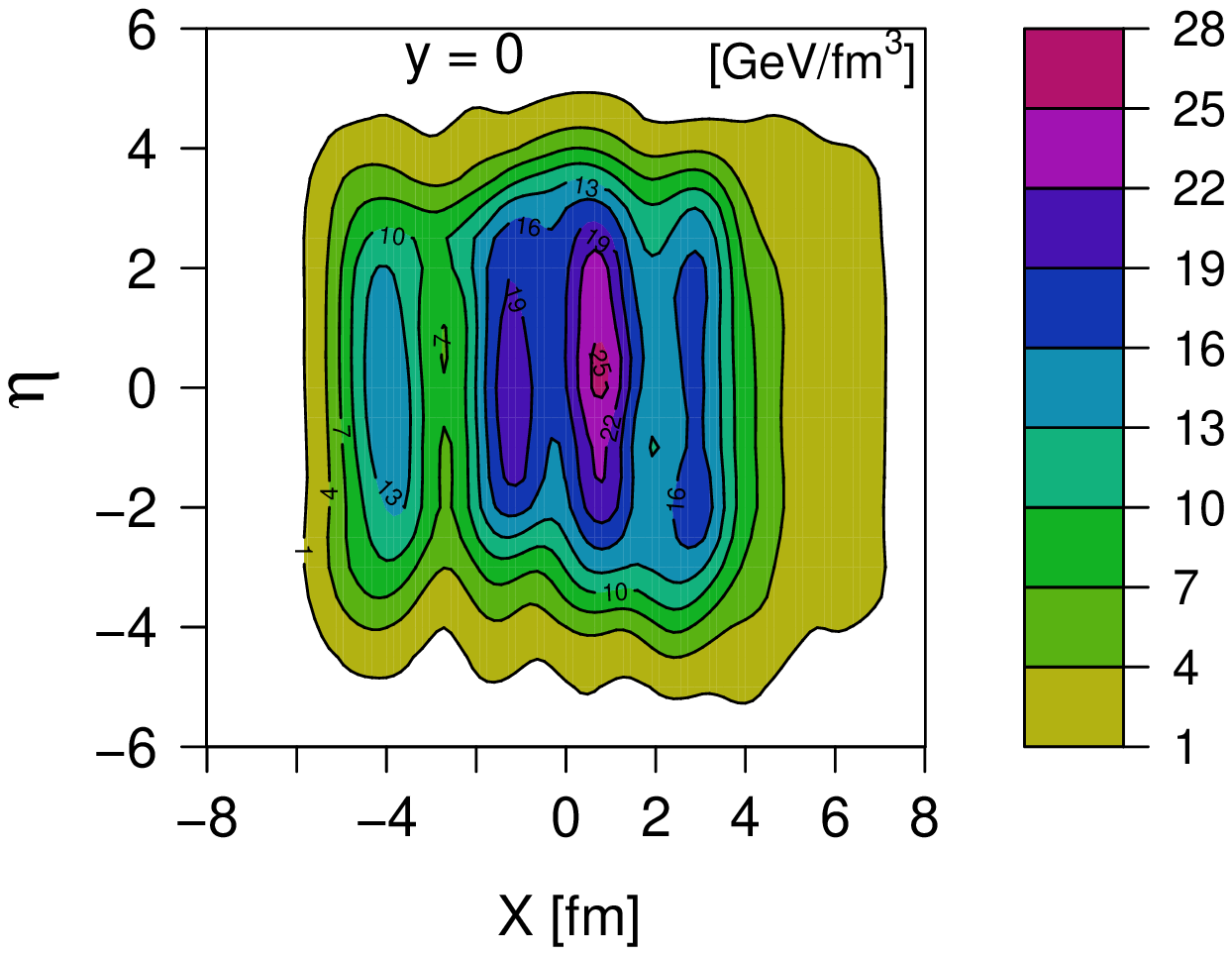} 
\includegraphics[width=6.cm]{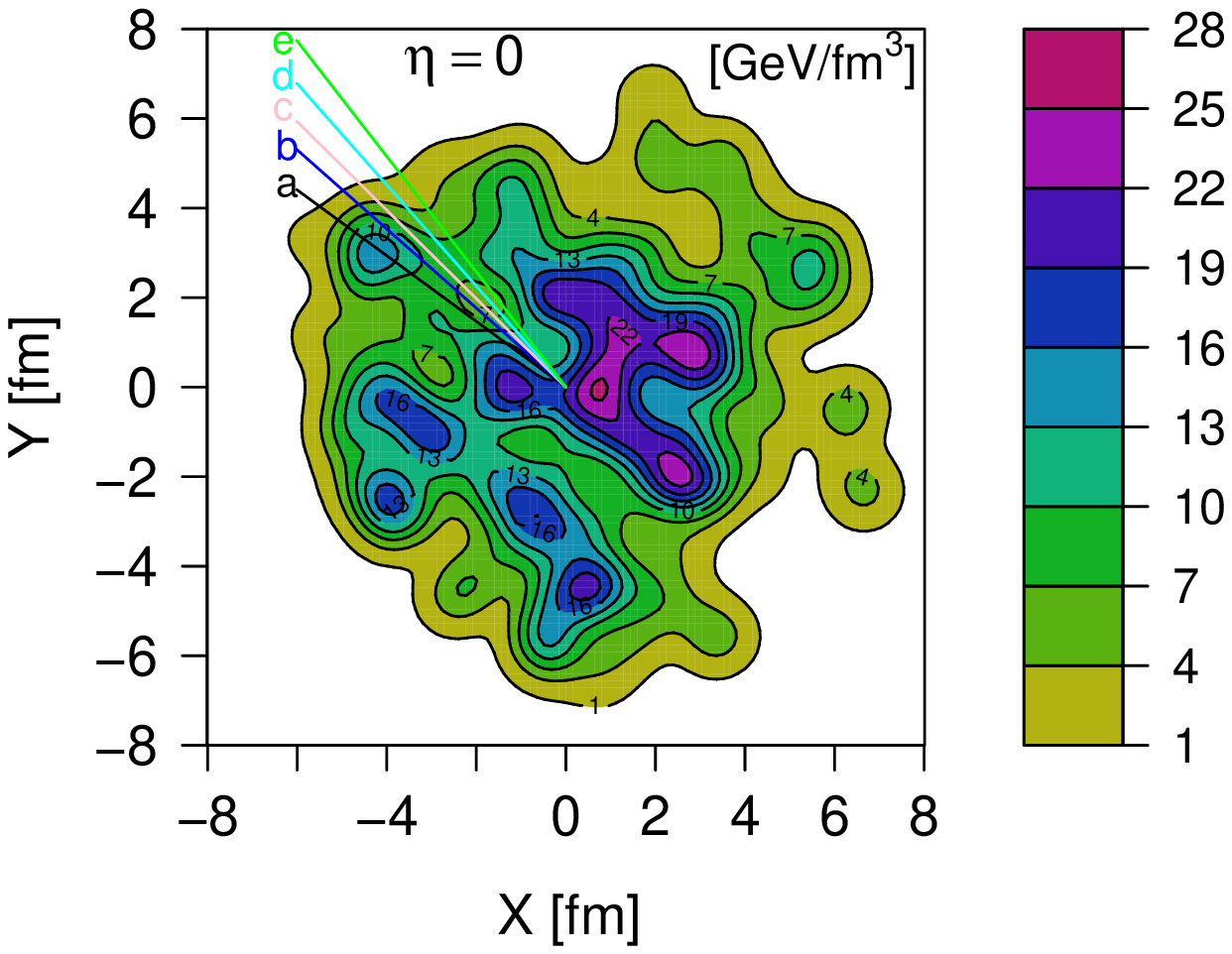}
\end{minipage} 
\begin{minipage}[t]{50mm} 
\vspace{-1.9cm} 
\hspace{-.5cm} 
\includegraphics[width=5.cm]{1c.eps} 
\end{minipage} 
\vspace{-.2cm} 
\caption{\label{ic} Left and center: initial energy density 
for a NeXus central Au-Au collision at 200 GeV A in the $y=0$ and  
$\eta=0$ plane respectively. 
Right: Comparison of the parametrization given by Eq.(1)  (solid lines) with the original NeXus energy density 
(dashed lines), along the lines a-e (in the $\eta=0$ plane). 
} 
\end{figure*}

We will consider central collisions only and use a 
simplified model. Fig. \ref{ic} (left and center) shows a 
typical example of initial conditions (initial energy 
density) obtained in NeXus with various tubular structures 
along the collision axis.  

The origin of these structures is the following.
To model soft physics in p-p collision, it is common to assume that strings
(or color flux tubes) are formed, either via the excitation of the protons or
due to color exchange between them. In A-A collisions, these strings may 
overlap leading to longitudinally extended regions of higher energy density
such as those in Fig. \ref{ic}. An alternative description of A-A collisions,
based on gluon saturation,
is that  the two colliding nuclei can be viewed as Color Glass Condensates.
Shortly after their collision, these produce strong color flux tubes called
``Glasma''. Therefore the possibility that tubular structures exist in the 
initial 
conditions is general but their exact characteristics are not known.

In the simplified model, only one 
of the high energy tubes from NeXus (chosen close 
to the border) is considered and the complex background is 
smoothed out. 
This leads to the following parametrization of the initial energy density 
\begin{equation} 
\epsilon=12 \exp\,(-0.0004r^5)+ \frac{34}{0.845 \pi}
 \exp\left(\frac{-|\vec r- \vec r_0|^2}{0.845}\right), 
\end{equation} 
where $r_0=5.4\,$fm.
A comparison of this parametrization with the original NeXus energy density is shown in Fig. \ref{ic} (right). 
Except for the inner region (which has little importance cf. \S 2.2), 
the agreement is reasonable. 
We use this parametrization in order to have a realistic tube description.
However as already mentioned, the exact characteristics of the color tubes
are not well known, therefore later we will consider various variations
of the parameters.

\begin{figure}[htb] 
\begin{center}
\vspace*{-0.7cm}
\includegraphics[width=5.1cm]{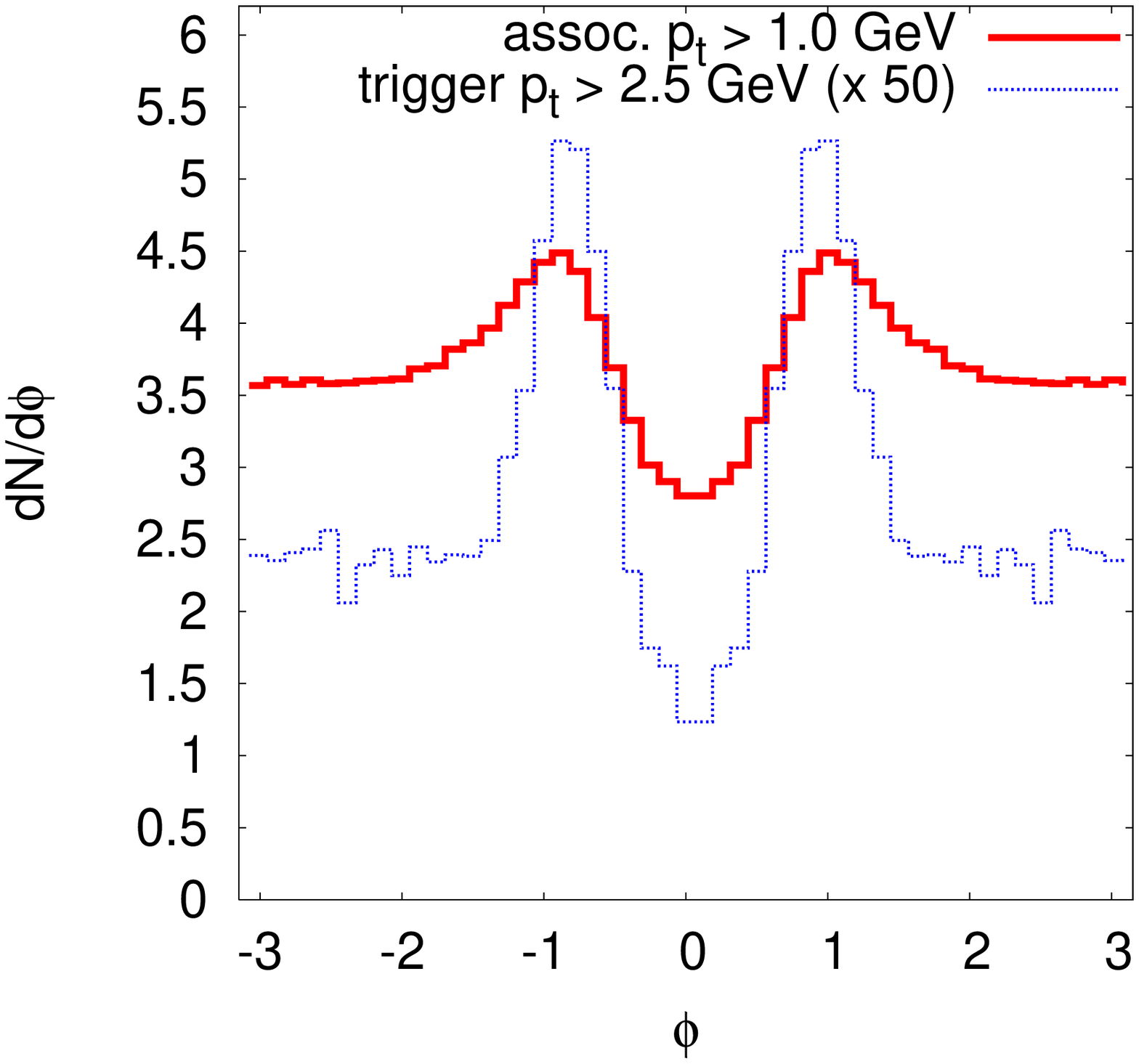} 
\includegraphics[width=5.3cm]{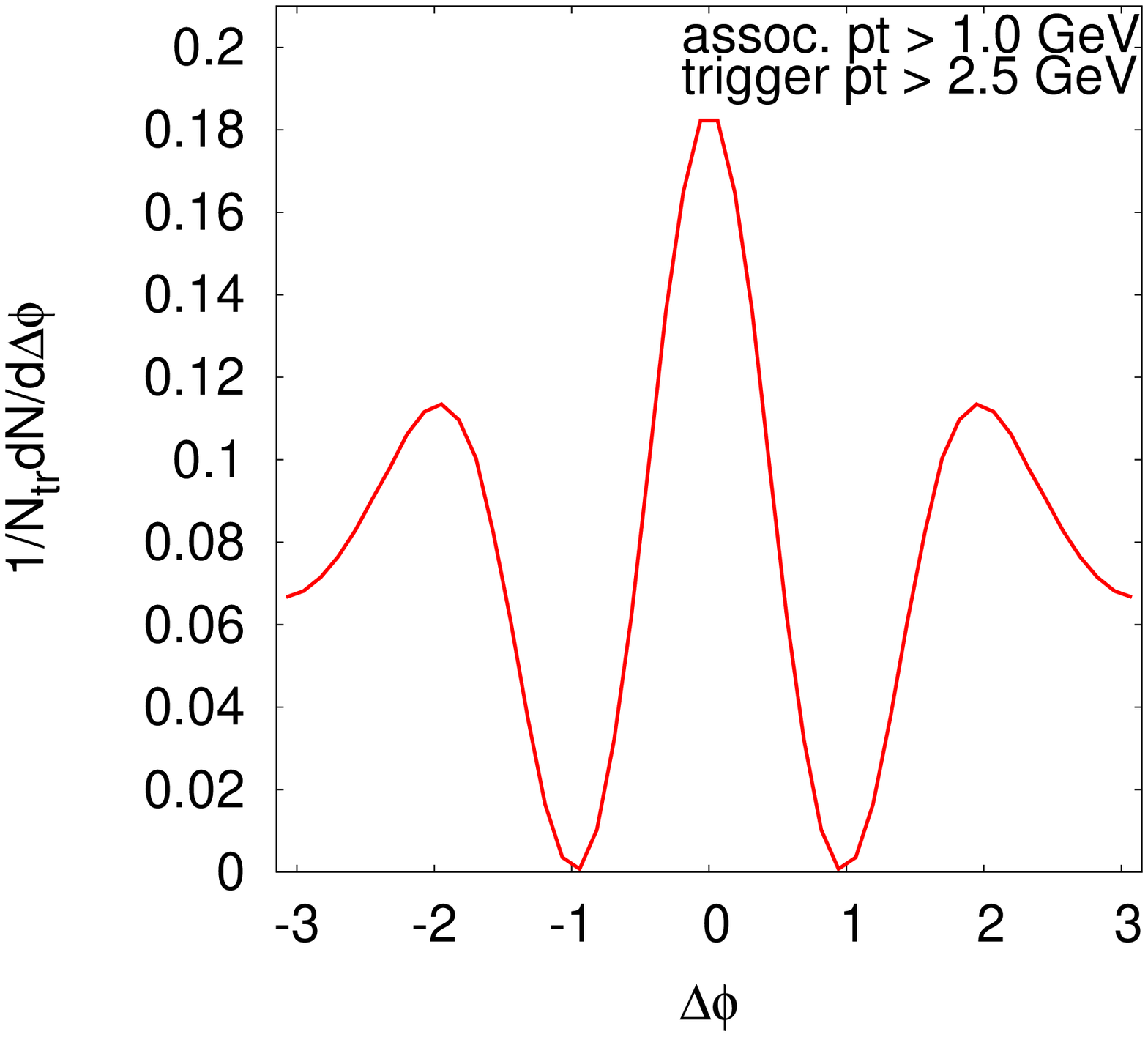} 
\vspace{-.2cm} 
\caption{\label{1dist} 
Angular distributions of (direct) charged particles
in some different $p_T$ 
intervals (top) and resulting two-particle correlations (bottom) in the simplified model (for a freeze out temperature of 0.14 GeV).} 
\end{center}
\end{figure}

In this simplified model or one-tube model,
transverse expansion is computed numerically while 
longitudinal expansion is assumed boost-invariant, until 
freeze out at some constant temperature. 
The resulting 
single-particle angular distribution,
shown in  Fig. \ref{1dist} (top),
has two peaks located 
on both sides of the position of the tube (placed at $\phi=0$) 
with separation $\sim 2$
(this is not a parameter),
more or less independently of the value of
the transverse momentum. Particle emission is computed assuming sudden freeze out. Since this is 
an approximation to real particle emission, we have checked that varying the 
freeze out temperature (between 130 and 150 MeV) does not affect qualitatively
our result.

\begin{figure*} 
\hspace*{-1.cm}
\includegraphics[width=6.1cm]{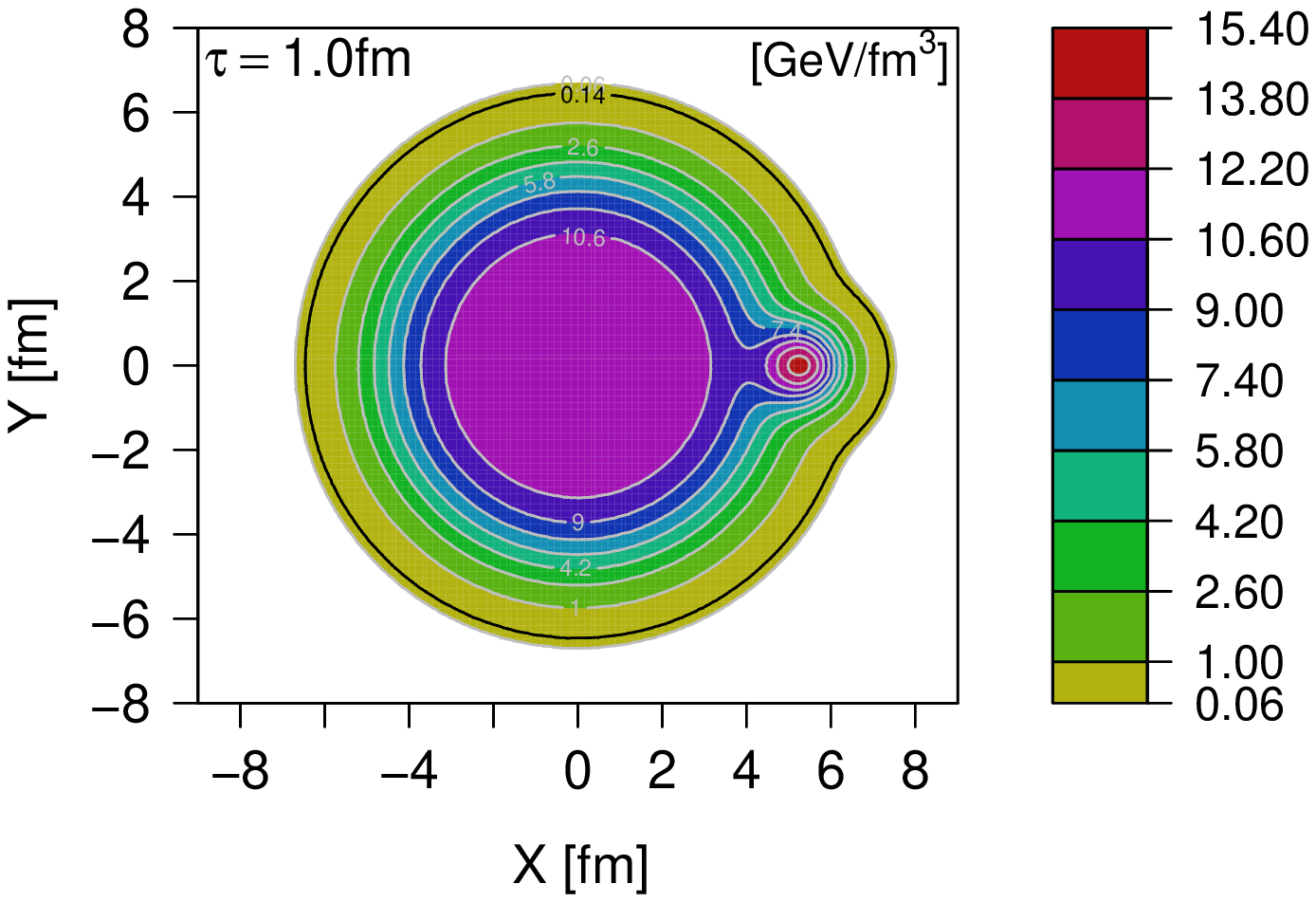}\includegraphics[width=6.1cm]{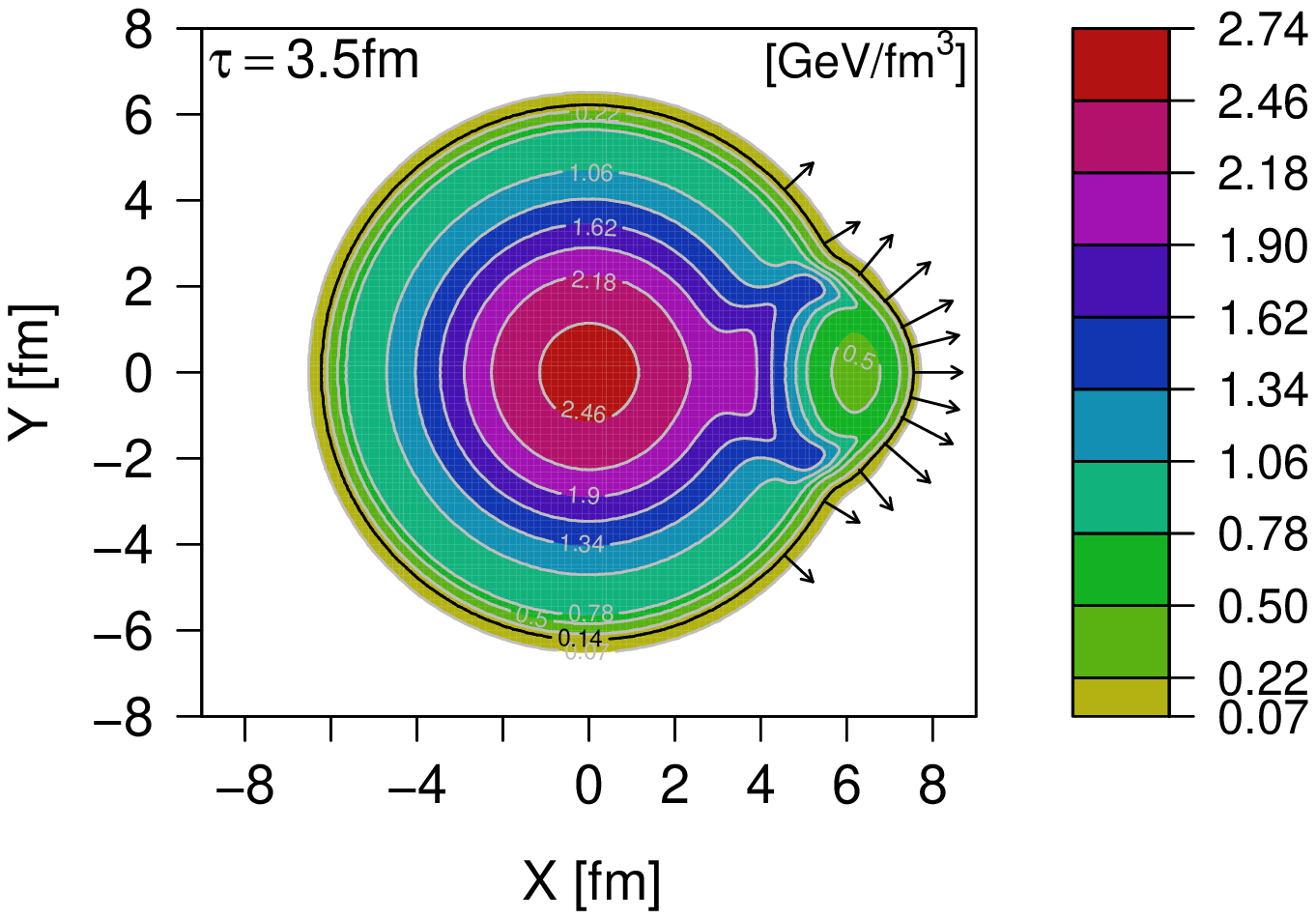}
\includegraphics[width=6.1cm]{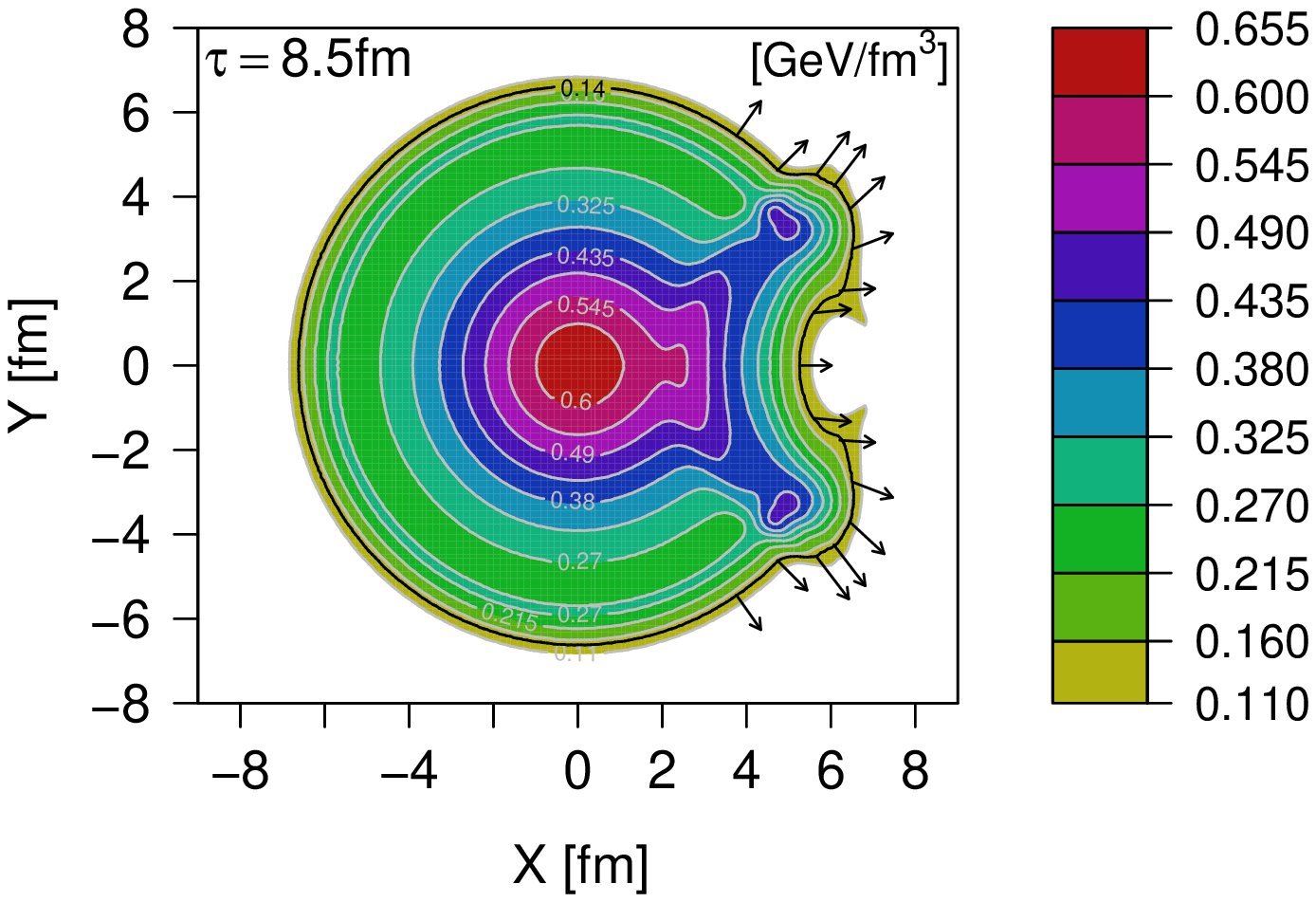} 
\vspace{-.2cm} 
\caption{\label{temp} Temporal evolution of energy density 
 for the simplified model at times 1.0, 3.5 and 8.5 fm
(thicker outer black curve corresponding to
 freeze out temperature of 0.14 GeV). 
Arrows indicate fluid velocity on the freeze out 
surface (vector length equivalent to 2 fm corresponds to light speed).} 
\end{figure*}

This two-peak emission is in contrast with what happens when 
a blast wave is assumed, namely the fact that high-energy 
tubes emit in a single direction. However, its occurrence 
can be understood from Fig. \ref{temp}. 
As time goes on, as 
a consequence of the tube expansion, a hole appears at the 
location of the high-energy tube (as in \cite{shuryakhole}). 
This hole is surrounded by matter that piles up in a roughly  semi-circular cliff of high-energy-density matter, guiding 
the flow of the background matter into two well-defined 
directions. 
The two extremities of the cliff emit more 
particles than the background, this gives rise to the 
two-peaks in the single-particle angular distribution. The 
emission is not quite radial as shown by 
Fig. \ref{temp} (right), indicating that there was a 
deflection of the background flow due to the pressure exerted
by the high-energy tube. 
As seen in  fig. 3, the fluid velocity is larger at the two extremities of the
cliff and smaller nearby, this is why in fig.2  
the angular distribution is narrower for larger $p_t$ particles.

From Fig. \ref{1dist} (top), we can guess how the 
two-particle angular correlation will be. The trigger 
particle is more likely to be in one of the two peaks. We 
first choose the left-hand side peak. The associated particle 
is more likely to be also in this peak i.e. with 
$\Delta\phi=0$ or in the right-hand side peak with 
$\Delta\phi\sim+2$. If we choose the trigger particle in the 
right-hand side peak, the associated particle is more likely 
to be also in this peak i.e. with $\Delta \phi=0$ or 
in the left-hand side peak with $\Delta \phi\sim -2$. 
So the final two particle angular correlation must have a 
large central peak at $\Delta\phi=0$ and two smaller peaks 
respectively at $\Delta\phi\sim\pm 2$. Fig. \ref{1dist} 
(bottom) shows that this is indeed the case.
The peak at $\Delta \phi=0$ corresponds to the near-side 
ridge and the peaks at $\Delta\phi\sim\pm 2$ form the 
double-hump ridge. 
We have checked \cite{Ronedoc}
that this structure is 
robust by studying the effect of the height ($12\pm 3$ in the first term
on the right hand side of eq. (1)) and shape of the 
background ($r^5$ to $r^2$ in the same term), overall 
initial transverse  velocity (increasing radially up to 0.6), 
height, radius and location of 
the tube (some details are shown in the next section).
The model was also
generalized to non-central collisions and the  
in-plane/out-of-plane trigger dependence studied \cite{Ronedoc}. 

\subsection{Dependence on the tube parameters}

\begin{figure}[htb]  
\begin{center}
\vspace{-.7cm} 
\includegraphics[width=5.2cm]{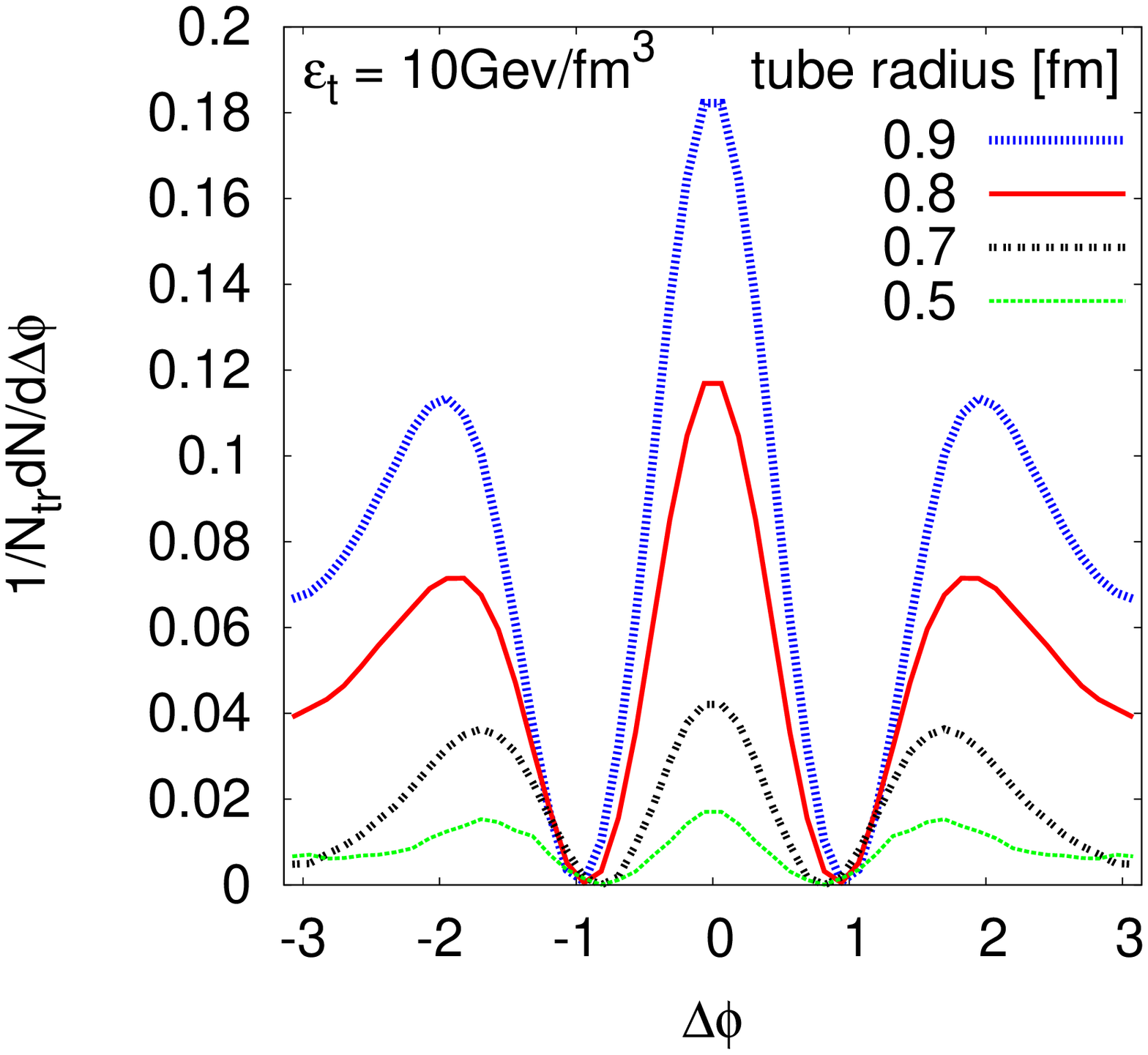}

\vspace*{-0.5cm}
\includegraphics[width=5.2cm]{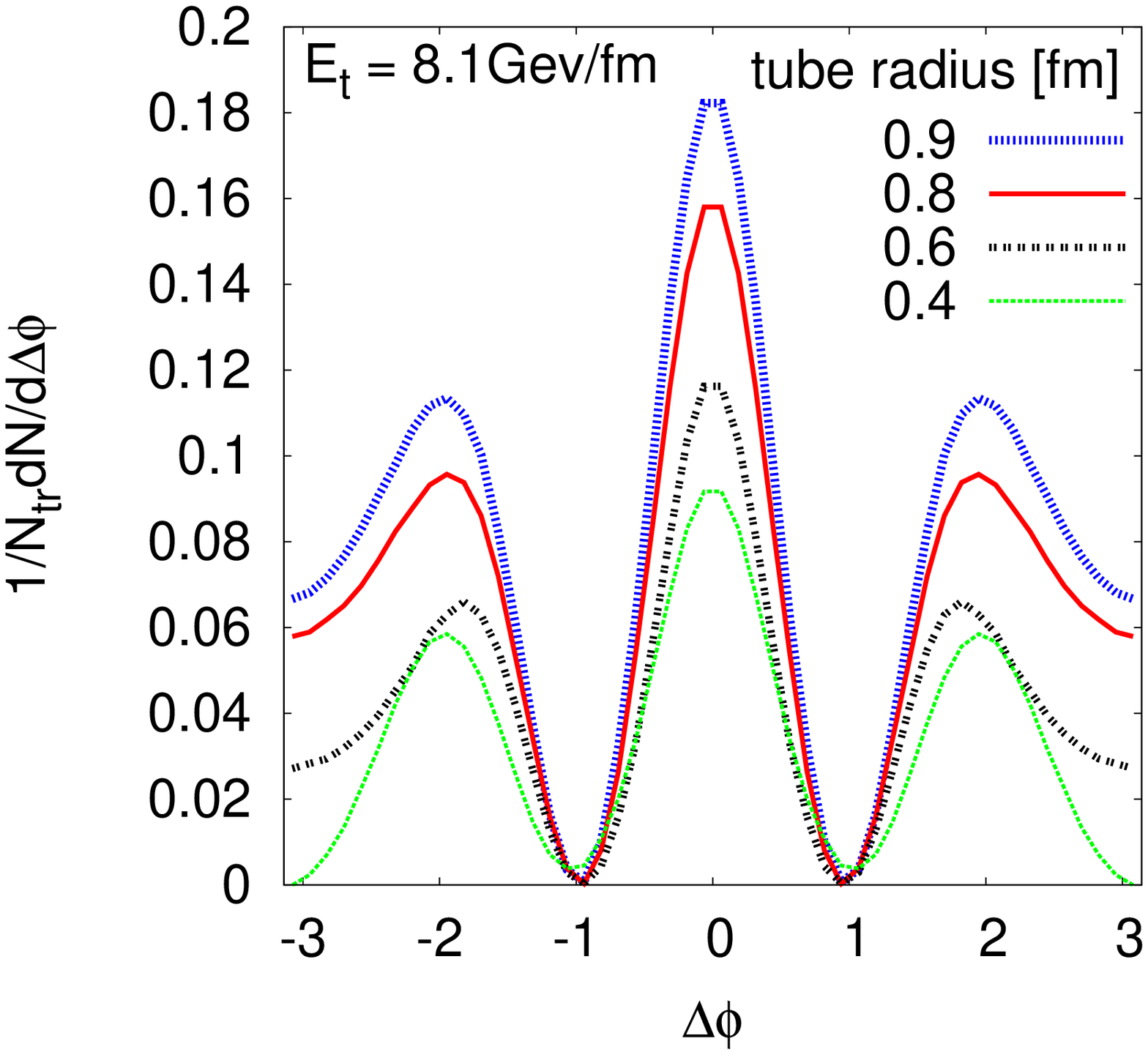}
\vspace{-.2cm} 
\caption{\label{raio} 
Top: two-particle correlation for tubes with different 
radius but similar energy density.
Bottom: two-particle correlation for 
tubes with different radius but similar energy content. 
(Tube position fixed at $r_0=5.4\,$fm.)}
\end{center}
\end{figure}

In the above calculation, the tube extracted from 
NeXus initial conditions  has a radius $\Delta r$ of order 0.9 fm 
and (maximum) energy density $\epsilon_t$ 
of order 12 GeV $fm^{-3}$ (at proper time 1 fm),
as can be seen from the second term on the right hand side of eq. (1) or 
from fig. 1.
Changing $\Delta r$  affects the height of the peaks and spacing in
the two-particle correlation as shown 
in Fig. \ref{raio} (top).
Changing $\epsilon_t$ has a similar strong effect.
On the other side, if $\Delta r$ and  $\epsilon_t$ are changed 
maintaining constant
the energy per unit length 
$E_t\propto\epsilon_t\,(\Delta r)^2$,
 as shown in Fig. \ref{raio} (bottom), 
the two-particle correlation maintains its overall shape, 
the angle between the peaks is almost unchanged and the peak 
heights change less. 
Therefore a good parameter to characterize the two-particle 
correlation is the tube total energy per unit length and  
tubes thinner than  0.9 fm are not excluded (see also 
\cite{klaus}). 
For this comparison the background was kept unchanged, so 
the thinner tube energy density is really much higher than 
the background one, more realistic cases could be studied 
but this goes beyond the scope of this paper.

\begin{figure}[t]
 \begin{center}
\vspace{-.7cm} 
\includegraphics[width=5.4cm]{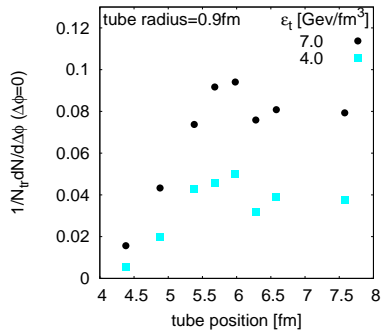}
\vspace*{-.2cm}
\caption{\label{location} Height of the two-particle correlation at $\Delta \phi=0$ as function of the tube position with respect to the center.} 
\end{center}
\end{figure}

The observation that the correlation is characterized by the energy per unit 
length
$E_t$ 
(not by $\Delta r$ or $\epsilon_t$ separately)
is consistent with the fact that what matters is the amount of energy available for the tube to push the surrounding matter.
This also explains why the precise shape of the tube energy density 
(e.g. gaussian as in eq. (1))
is not  crucial.

Finally, in Fig. \ref{location},
 the
tube is located (above the background) at various distances 
$r_0$
from the 
center, 
the height of the two-particle correlations at $\Delta \phi=0$ 
saturates between 
5.5 and 8 fm and decreases strongly for smaller distances to 
the center.

The physical reason for the behavior of the maximum height of the two-particle correlation as function of the tube position
 can be understood by looking at the temporal evolution of the energy density. 
When the tube is exactly at the center, there is no 
privileged direction of emission. When the tube is close to the outer border 
two privileged directions of emission appear (cf. Fig. \ref{temp}). When the
tube is at some small distance such as 2 fm, even though the 
strong expansion of the
tube presses the surrounding matter creating a 
hole, this happens  too much inside to cause the appearance of the 
two privileged directions of emission.
We conclude, as mentioned earlier, that only peripheral tubes are 
important for the particle correlation\footnote{
An opposite conclusion
was reached in  fig. 10 of \cite{shuryakG}.
A possible explanation for this discrepancy, is that they
 use  the Gubser solution which has some unphysical features at large radii
as discussed in  \cite{gubser} (in particular see fig. 3 there).}.

\section{More realistic case}

\begin{figure*}[t]  
\begin{minipage}[h]{130mm}
\hspace*{-2.3cm} 
\includegraphics[width=9.6cm]{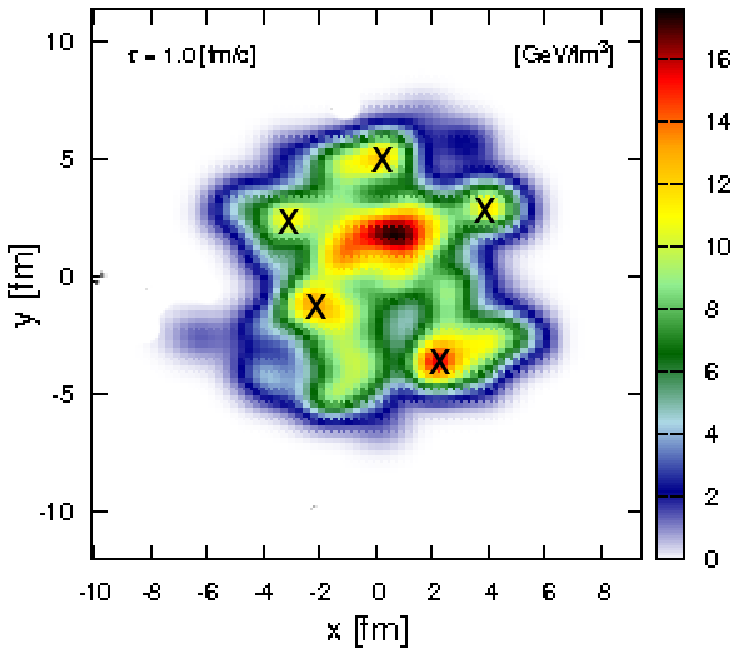} 
\hspace*{-2.cm}\includegraphics[width=9.6cm]{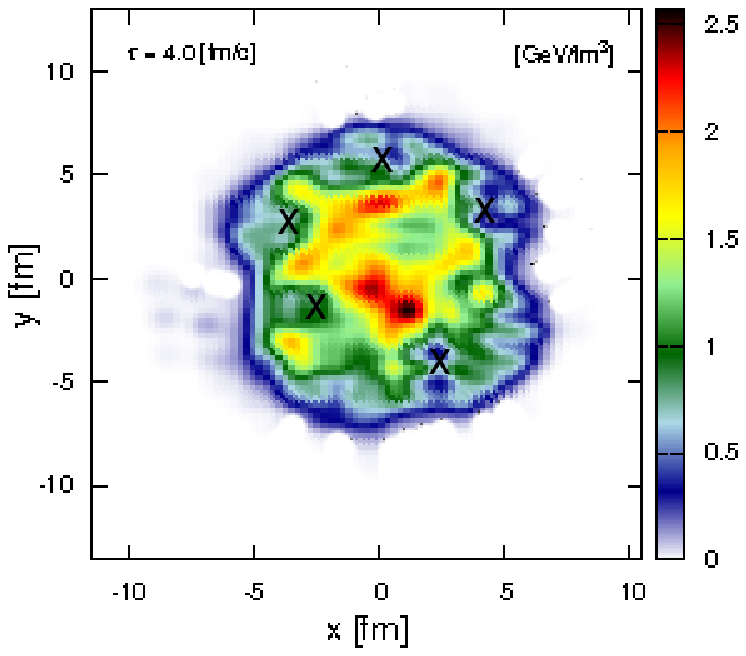}
\end{minipage} 
\begin{minipage}[t]{50mm} 
\vspace{-3.6cm} 
\includegraphics[width=6.2cm]{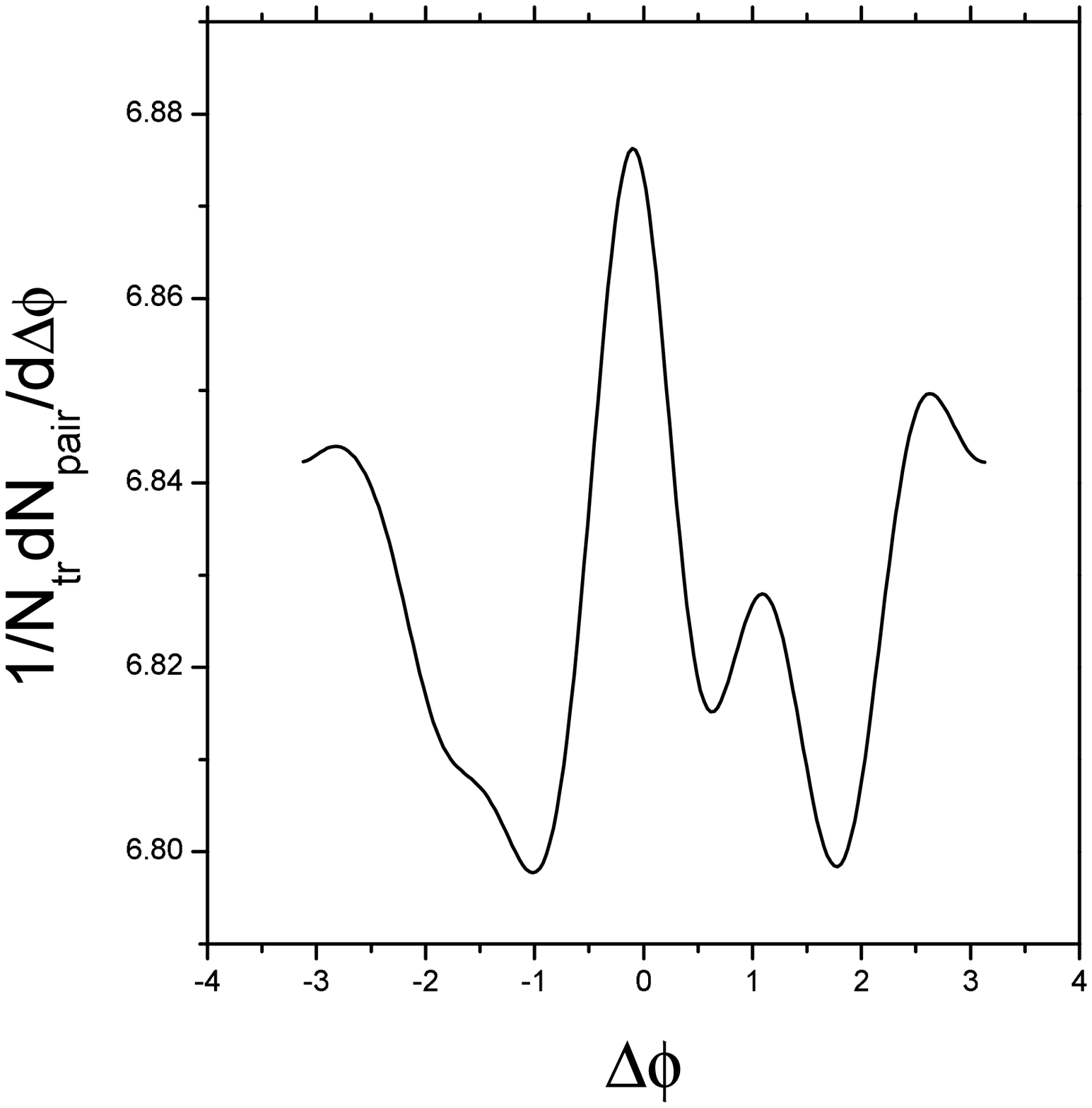} 
\end{minipage} 
\vspace{-2.1cm} 
\caption{\label{nex6} Left: initial energy density 
for a NeXus central Au-Au collision at 200 GeV A in the   
$\eta=0$ plane with five  outer tubes.
Center: appearance of holes in the energy density
during the temporal evolution (obtained assuming longitudinal boost invariance).
Right: two-particle correlations. 
} 
\end{figure*}

With these information (and using the same two-dimensional hydrodynamic model 
as in previous section), 
 we can discuss what happens in a 
more complex event such as a NeXus event. In such an event, 
only the outer tubes are expected to be relevant, 
for example in  fig. \ref{nex6} (left), we can pinpoint five such tubes,
indicated by crosses.
When the time evolution of this matter slice is studied, holes appear
in the vicinity of the
 former location of the tubes indicated by crosses in fig. \ref{nex6} (center).
Due to expansion and the fact that one tube can interfere with another.
we do not expect perfect one-to-one correspondence 
(though in this particular event, it is approximately the case).

The shape of the two-particle correlations for a single tube  
(in particular the peak spacing) is relatively independent 
of its features so the various tubes in the NeXus event under study 
will contribute with 
rather similar two-peak emission pattern at various angles 
in the single-particle angular distribution. As a consequence,
the two-particle correlation of this NeXus event is expected to
have a well-defined main structure similar to that of the one-tube
model of the previous section 
(Fig. \ref{1dist}) surrounded by several other peaks and 
depressions due to trigger and associated particles coming 
from different tubes.  This is indeed the case as shown in Fig. \ref{nex6} 
(right).
The additional peaks and depressions 
have positions depending on the angle of the tubes between 
them.
When averaging over many events, these interference 
terms  cancel out and only the main one-tube-like 
structure is left \cite{jun}. 
In other words: the picture derived in section 2 also 
applies to more complex events such as Nexus ones. 


\section{Conclusion and perspectives}

Usually, the initial conditions in the hydrodynamic 
description of relativistic nuclear collisions are assumed 
to be smooth.
It seems however that each time more, understanding data 
requires a knowledge of the fluctuating event-by-event 
initial conditions rather than an assessment of some adequate
smooth initial conditions:
 fluctuations in elliptic flow \cite{spheriofluct} (perhaps the very behavior of elliptic 
flow as function of pseudorapidity \cite{spheriov2}),
Fourier coefficients of the azimuthal anisotropic flow (see below), 
two-particle correlations (see introduction).

In this paper, a unified picture for the  structures 
observed in two particle correlations at low to moderate 
transverse momentum has been presented. It is based on the 
presence of longitudinal high energy density tubes in the 
initial conditions. These tubes are leftover from the 
initial particle interactions. During the hydrodynamic 
evolution of the fluid, 
 the strong expansion of 
the tubes located close to the border piles up matter 
in two symmetrical 
directions, leading to  two-particle correlations with 
a near-side and a double hump away-side ridges (for central collisions). 

An alternative unified picture has also been suggested, the idea is the 
following \cite{alver}. The   ellipticity of the interaction region in a 
collision gives rise to elliptic flow because of the larger pressure 
gradients along the minor axis of the ellipse. 
Similarly, if the interaction region has  some triangular shape,
this causes triangular flow (due to existing larger pressure gradients in certain directions).
Both near-side and  double hump away-side ridges
 are a natural consequence of triangularity and 
triangular flow \cite{alver}.

More generally, it has been suggested that the initial transverse  density 
in the overlap region can be decomposed using an infinite set of moments and any observable (in particular
anisotropic flow Fourier coefficients $v_n$)  can be written as a function of these
moments \cite{Teaney}. Conventional eccentricities such as ellipticity and triangularity, 
are basically  a subset of these moments, which may or not be sufficient to characterize 
the $v_n$'s (see e.g. \cite{Gardim12}).

Both tube configuration  such as in this paper and triangularity \cite{alver} (or more 
general geometrical shape of the interaction region)
  lead to non-zero eccentricities. 
In the case of NeXus, these eccentricities reflect the
initial conditions where
a lot of peaks or ``hot spots'' and valleys are present  as can be seen in
 Fig. 1 and 6
(similar features  can be seen in EPOS
initial conditions \cite{klaus}). 
On the other side,
depending on the sampling process for the nucleon-nucleon collisions
and sources of fluctuations included, the
interaction region may present a geometrical shape without dominance of a few 
tubes  and might be  described by
conventional eccentricities such as suggested in \cite{alver}. 

The hydrodynamical response to the anisotropies differs in the two cases.
In our description, the structures in the two-particle correlations and
the various $v_n$'s
are a response to individual {\em outer} tubes: this is a local effect.
In the case of triangularity (or more general geometrical shape) of the interaction region, the
structures in the two-particle correlations and  $v_n$'s
correspond  
to the various geometrical 
deformations: this is a global effect. 
In our approach, the angular size of the near-side ridge,
$\sim 2$,
is of local hydrodynamic origin and can be roughly understood as 
the diameter of the hole divided by the radius of the nucleus, i.e.
$2 \,c_s\, \tau_{f.out}/R$ (ignoring the non-radial flow of particle
emission cf. \S 2.1).
In the triangularity case, this
 angular size is fixed by global geometry (for example $2 \pi /3$ 
in the simple case of an  equilateral
triangular shape).
Similarly, in our approach, the relative height of the peaks in the
two-particle correlation is of local hydrodynamical origin: for example, for a 
single tube (in a central collision), the highest peak is approximately twice 
higher than the two smaller peaks
(cf. fig.2). In the triangularity case,  the relative height of the peaks in
 the
two-particle correlation is related to global geometry: for example, in the 
simple
 case of an  equilateral
triangular shape, there should be three equal height peaks in the
two-particle correlation.
However, to turn the connection between initial (local or global)
geometry and flow more precise, 
pre-equilibrium evolution \cite{qin} and viscosity
\cite{schenke} should be included since both, though they are small effects,
can smear out the initial
anisotropies.

Finally, to further test 
the presence of  tubular structures in the initial conditions,
we suggest 
to build 2+1 correlations, fixing both a trigger
 particle and a first associated particle, this last one 
 with $\Delta\phi_1\sim 2$. 
This choice ensures that both particles come from different
emission peaks in the single particle angular distribution.
Then, in our approach, 
the second associated particle will be more likely to come
from the same emission peak as the trigger, i.e.
with $\Delta\phi_2=0$ or the same emission peak as the first associated
particle,  i.e.
with $\Delta\phi_2 \sim 2$.
Naively, the plot of the 2+1 correlation as function of 
$\Delta\phi_2$ vs. $\Delta\eta_2$ should present two stripes located at
$\Delta\phi_2=0$ and $\Delta\phi_2 \sim 2$.
However, in practice, there appears a third weaker stripe, due to the 
background. In the eccentricity case,  
the plot is expected to be more complicated
(in the simple case of
only equilateral triangular shapes - with longitudinal extension, 
there should be
three equally bright stripes). 
Additional work with realistic initial conditions
is needed to check whether  2+1 correlations may indeed  permit to 
distinguish between both scenarios. 

Since the two scenarios,  isolated
tube configuration
such as in this paper and triangularity  (or more 
general geometrical shape of the interaction region)
correspond to different schemes of
initial state energy deposition,
distinguishing between them may allow to learn about how the strong interaction
proceeds during
high energy nuclear collisions.

We acknowledge funding from 
 CNPq and FAPESP.





\bibliographystyle{unsrt}
\bibliography{ridge}







\end{document}